\pdfoutput=1
\documentclass[sigconf, screen]{acmart}

\renewcommand\footnotetextcopyrightpermission[1]{} %

\usepackage{todonotes}

\undef\definition
\undef\example
\newtheorem{definition}{Definition}
\newtheorem{example}{Example}

\usepackage{physics}
\usepackage{flushend}
\usepackage{nicematrix}
\usepackage{xfrac}
\usepackage{subcaption}
\usepackage{graphicx}
\usepackage{tikz}
\usepackage{pgfplots}
\usepackage{csvsimple}
\usetikzlibrary{shadows,arrows.meta,backgrounds,calc,chains,matrix,positioning,shapes}
\usetikzlibrary{arrows.meta,backgrounds,calc,chains,matrix,positioning,shapes,shapes.geometric,	shapes.arrows, decorations.pathmorphing, decorations.pathreplacing}
\tikzset{%
	font={\footnotesize},
	vertex/.style={draw,circle,inner sep=0pt,minimum width=0.5cm,minimum height=0.5cm,font=\small, scale=0.9},
	terminal/.style={draw,regular polygon,regular polygon sides=4,inner sep=0pt,minimum width=0.5cm,minimum height=0.5cm,font=\small, scale=1.0},
	zeroterm/.style={below,inner sep=0pt,font=\scriptsize, scale=0.9}
}
\pgfplotsset{compat=1.15}
\pgfkeys{/pgf/number format/.cd,1000 sep={\,}}

\setcopyright{acmcopyright}
\copyrightyear{2020}
\acmYear{2020}
\setcopyright{acmcopyright}
\acmConference[ICCAD '20]{IEEE/ACM International
Conference on Computer-Aided Design}{November 2--5, 2020}{Virtual Event, USA}
\acmBooktitle{IEEE/ACM International Conference on Computer-Aided Design (ICCAD'20), November 2--5, 2020, Virtual Event, USA}
\acmPrice{15.00}
\acmDOI{10.1145/3400302.3415622}
\acmISBN{978-1-4503-8026-3/20/11}

\begin{document}

\title{Considering Decoherence Errors in the\\Simulation of Quantum Circuits Using Decision Diagrams} %

\author{Thomas Grurl$^{*\dagger}$,\hspace{1em} Jürgen Fuß$^{*}$,\hspace{1em} Robert Wille$^{\dagger\ddagger}$ }
\affiliation{%
 \institution{
 \mbox{\hspace{13mm}}$^{*}$Secure Information Systems, University of Applied Sciences Upper Austria, Austria\newline 
 }\vspace{-4mm}
}
\affiliation{%
 \institution{
 \mbox{\hspace{14mm}}$^{\dagger}$Institute for Integrated Circuits, Johannes Kepler University Linz, Austria\newline
 }\vspace{-4mm}
}
\affiliation{%
 \institution{
 \mbox{\hspace{13mm}}$^{\ddagger}$Software Competence Center Hagenberg GmbH (SCCH), Hagenberg, Austria\newline
 }\vspace{-4mm}
}
\affiliation{
 \institution{
 {\hspace{-4mm}\{thomas.grurl, juergen.fuss\}@fh-hagenberg.at,\hspace{1em}robert.wille@jku.at}
 }
}\vspace{-4mm}
\affiliation{
 \institution{
 {{http://iic.jku.at/eda/research/quantum/}}
 }
}

\begin{abstract}
By using quantum mechanical effects, quantum computers promise significant speedups in solving problems intractable for conventional computers. However, despite recent progress they remain limited in scaling and availability---making quantum software and hardware development heavily reliant on quantum simulators running on conventional hardware. However, most of those simulators mimic \emph{perfect} quantum computers and, hence, ignore %
the fragile nature of quantum mechanical effects which frequently yield to decoherence errors in real quantum devices.
Considering those errors during the simulation is complex, but necessary in order to tailor quantum algorithms for specific devices.
Thus far, most state-of-the-art simulators considering decoherence errors rely on (exponentially) large array representations. 
As an alternative, simulators based on decision diagrams have been shown very promising for simulation of quantum circuits in general, but have not supported decoherence errors yet. 
In this work, we are closing this gap. We investigate how the consideration of decoherence errors affects the simulation performance of approaches based on decision diagrams and propose advanced solutions to mitigate negative effects. Experiments confirm that this yields improvements of several orders of magnitudes compared to a naive consideration of errors.

\end{abstract}

\maketitle
\renewcommand{\shortauthors}{Thomas Grurl, Jürgen Fuß, Robert Wille} 
\section{Introduction}

Quantum computers can solve specific problems significantly faster than classical computers. Besides early examples of corresponding algorithms such as Shor's factorization approach~\cite{DBLP:journals/siamcomp/Shor97} and Grover's database search~\cite{DBLP:conf/stoc/Grover96}, recently also other relevant quantum algorithms have been found in the areas of chemistry, finance, machine learning, and mathematics~\cite{montanaro2016quantum,Kassal18681, RebentrostFinance,KerendisQmeans, farhiQAOA}. 
Moreover, not least due to the drive of big companies such as  IBM, Google, Intel, Rigetti, Microsoft, or Alibaba, who are heavily investing in this emerging technology, there have been remarkable accomplishments towards the physical realization of quantum computers recently.
\newpage
However, quantum computers do not work perfectly and are indeed even more prone to errors than their classical counterparts. 
In particular, decoherence errors~\cite{Tannu2018} are a common phenomenon with which researchers and engineers frequently have to deal with. 
They lead to the problem that qubits can only hold information for a limited amount of time. Although 
recent developments in the physical realization of quantum computers have improved upon that~\cite{Devoret1169, Kelly2018}, decoherence errors are still a dominating aspect in quantum computing. 

Accordingly, it is essential to specifically evaluate whether quantum algorithms can cope with this kind of errors. To this end, methods for quantum circuit simulation would be ideal as they allow for an explicit investigation how a given algorithm behaves on a specific device and its possible errors.
But most of the available simulators (such as, e.g., proposed in~\cite{vidal2003efficient,zulehner2017advanced,MTG:2006,niemannQMDDsEfficientQuantum2016,Steiger2018projectqopensource}) mimic \emph{perfect} quantum computers and, thus, ignore the fragile nature of quantum mechanical effects which frequently yield to decoherence errors.
Luckily, the effects are well understood and mathematical models for decoherence errors are available~\cite{NC:2000}---leading to first simulation approaches to also support the consideration of those errors (e.g. ~\cite{qiskit,atos2016,qxSimulator2017,DBLP:journals/corr/WeckerS14,CirqPythonFramework,jones2018quest,DBLP:journals/corr/SmelyanskiySA16,villalonga2019highperformancesimulator}).

But a main challenge remains: The capabilities of those simulation approaches is severely limited  by the inherent exponential nature of the vectors and matrices which describe the respective quantum states and operations, respectively. Additionally considering decoherence errors further increases the resulting complexity. 
Because of this, alternatives to straightforward approaches (which represent the required vectors and matrices in terms of arrays) are currently investigated.
Decision diagrams for quantum circuit simulation %
(as introduced, e.g., in~\cite{DBLP:conf/esa/Samoladas08,viamontes2004high,zulehner2017advanced,MTG:2006,niemannQMDDsEfficientQuantum2016}) provide a promising approach. In many cases and/or for many quantum applications, they allow for a representation of vectors and matrices which is below the exponential size of array-based solutions~\cite{zulehner2017advanced,grurl2020caseStudy}.

All these methods, however, do not support the consideration of decoherence errors yet. Moreover, thus far, it remains unknown whether the promising effects of a more efficient representation of vectors and matrices can still be maintained, when those errors are additionally taken into account. 
In this work, we are investigating this issue. 
Our observations and evaluations suggest that decision diagrams might remain compact in many cases---even if decoherence errors are considered.
But we also show that just having a compact representation is not sufficient: Efficiently realizing the operations describing the error effects poses a substantial challenge.
In order to mitigate those negative effects, we eventually propose an advanced solution for simulation using decision diagrams that also can efficiently handle decoherence errors. 

Experimental evaluations and comparisons to state-of-the-art simulators by IBM and Atos confirm %
the viability of the proposed solution.
Moreover, we show that the proposed advanced solution
is able to complete simulation runs several orders of magnitudes faster than the solution with a naive consideration of errors.

Our contributions are described in the rest of this paper as follows: Section~\ref{sec:background} reviews quantum computing and quantum circuit simulation using arrays as well as decision diagrams. Section~\ref{sec:sim} discusses how considering errors changes the way simulation is conducted and analyzes the resulting challenges for approaches based on decision diagrams. 
Based on these insights, we propose an advanced simulation scheme in Section~\ref{sec:opt}. Finally, we summarize our evaluations in Section~\ref{sec:eval} and conclude the paper in Section~\ref{sec:conclusion}. 

\section{Background}
\label{sec:background}
In this section, we review the basics of quantum computing and the simulation of corresponding quantum algorithms/circuits (using array-based methods as well as methods based on decision diagrams).

\subsection{Quantum Computing}
\label{sec:bg_quantum}
While, in the classical world, systems are described by bits which can either be 0 or 1, in quantum computing a system is described by \mbox{so-called} quantum bits or qubits. In contrast to classical bits, qubits can assume not only the state 0 or 1---which are called basis states and, using Dirac notation, are written as $\ket{0}$ and $\ket{1}$---but also an almost arbitrary combination (\emph{superposition}) of these states. %
More precisely, the state of the qubit $\ket{\phi}$ is given by $\ket{\phi} = \alpha_0 \cdot \ket{0} + \alpha_1 \cdot \ket{1}$. The amplitudes $\alpha_0, \alpha_1 \in \mathbb{C}$ describe how the qubit is related to each of the basis state and must satisfy the normalization condition $\abs{\alpha_0}^2 + \abs{\alpha_1}^2 = 1$. Measuring a qubit in superposition results in its collapse into one of the basis states $\ket{0}$ or $\ket{1}$ with probability $\abs{\alpha_0}^2$ and $\abs{\alpha_1}^2$, respectively.

These concepts can be extended for \mbox{multi-qubit} systems---to \mbox{so-called} \emph{quantum registers}---to represent the exponential number of basis states the system can assume. For example, a two qubit system~$\ket{\psi}$ has four basis states and is described by $\ket{\psi} = \alpha_{00} \cdot \ket{00} + \alpha_{01} \cdot \ket{01} + \alpha_{10} \cdot \ket{10} + \alpha_{11} \cdot \ket{11}$. Usually, the state description for \mbox{n-qubit} systems is shortened to a column vector of size $2^n$ containing only the amplitudes, e.g., $ \begin{bNiceMatrix}[small] \alpha_{00}&\alpha_{01}&\alpha_{10}&\alpha_{11}\end{bNiceMatrix}^\top$ for $n=2$.

The state of a qubit can be manipulated using quantum operations. With the exception of the measurement operation, all quantum operations are inherently reversible and, therefore, represented by unitary matrices. Important quantum operations are the %
NOT operation, which negates the state of a qubit, and the Hadamard operation, %
 which transforms a qubit from a basis state into a superposition. In addition to single-qubit operations, there are two-qubit operations. A prominent operation is the controlled-NOT (CNOT) operation, which negates the state of a qubit, if the designated control qubit is in state $\ket{1}$. 
Applying an operation to a quantum state is done via matrix-vector multiplication. 
To illustrate these concepts, consider the following example:
\begin{example}
\label{exp:matrxi_vector_mul}
Consider a two-qubit register in the state 
\begin{align*}
\ket{\psi} = \frac{1}{\sqrt{2}} \cdot \ket{00} + 0 \cdot \ket{01} + \frac{1}{\sqrt{2}} \cdot \ket{10} + 0 \cdot \ket{11},
\end{align*}
which is represented as $ \sfrac{1}{\sqrt{2}}\cdot\begin{bNiceMatrix}[small] 1&0&1&0\end{bNiceMatrix}^\top$. This is a valid state since $\abs{{\sfrac{1}{\sqrt{2}}}}^2 + 0^2+ \abs{{\sfrac{1}{\sqrt{2}}}}^2 + 0^2 = 1$. Measuring the state yields either $ \ket{00} $ or $ \ket{10} $, both with probability $\abs{{\sfrac{1}{\sqrt{2}}}}^2 = {\sfrac{1}{2}}$. 
Applying a CNOT operation to the state, which flips the amplitude of the second qubit when the first qubit, is set to 1 is given by
\begin{align*}
\underbrace{\begin{bmatrix}1&0&0&0\\0&1&0&0\\0&0&0&1\\0&0&1&0\end{bmatrix}}_{\textup{CNOT}}  \cdot  \underbrace{\frac{1}{\sqrt{2}}\begin{bmatrix}1\\0\\1\\0\end{bmatrix}}_{\mathit{\ket{\psi}}} = \underbrace{\frac{1}{\sqrt{2}}\begin{bmatrix}1\\0\\0\\1\end{bmatrix}}_{\mathit{\ket{\psi^\prime}}}.
\end{align*}

Measuring the new state $\ket{\psi^\prime}$ now yields either $ \ket{00} $ or $ \ket{11} $, each with probability ${\sfrac{1}{2}}$.

\end{example}

\subsection{Array-based Quantum Circuit Simulation}
\label{sec:array_based_sim}
The array-based simulation style realizes the concepts described above in a straightforward fashion. States and operations are represented by 1-dimensional and 2-dimensional arrays, respectively. Simulation is conducted by matrix-vector multiplications similar to Example~\ref{exp:matrxi_vector_mul}. Since the multiplication of a matrix and a vector can be decomposed into smaller operations, this simulation approach has a huge potential for parallelization. Every matrix-vector multiplication can be split into a series of multiplications and additions, i.e.,
\begin{align*}
\begin{bmatrix}
M_{00} & M_{01} \\
M_{10} & M_{11} \\
\end{bmatrix}
\cdot
\begin{bmatrix}
V_{0} \\
V_{1} \\
\end{bmatrix}
=
\begin{bmatrix}
	M_{00} \cdot V_0 + M_{01} \cdot V_1 \\
M_{10} \cdot V_0 + M_{11} \cdot V_1 \\
\end{bmatrix}.
\end{align*}
This can be further decomposed in a recursive fashion, leading to a large set of intermediate operations, which can be executed independently of each other with little synchronization overhead. State-of-the-art array-based simulators (such as~\cite{DBLP:journals/corr/WeckerS14,DBLP:journals/corr/SmelyanskiySA16,qxSimulator2017,qiskit,Steiger2018projectqopensource,CirqPythonFramework,atos2016}) make heavy use of that. 

\subsection{Decision Diagram-based Quantum Circuit Simulation}
\label{subsec:dd_simulation}

In order to tackle the memory problem of array-based simulators, a complementary approach has been developed using decision diagrams~\cite{DBLP:conf/esa/Samoladas08,viamontes2004high,zulehner2017advanced,MTG:2006,niemannQMDDsEfficientQuantum2016}. The general idea of decision \mbox{diagram-based} simulation is about identifying data redundancies in the state and representing them using shared sub-structures. Doing so results in a potentially very compact state representation, which in turn allows simulating quantum applications that cannot be simulated using array-based simulation approaches. 

Representing a state vector as a decision diagram revolves around recursively splitting the vector into equal sized sub-vectors, until the sub-vectors only contain a single element. More precisely, consider a quantum register $q_0, q_1,\dots,q_{n-1}$ composed of $n$ qubits, where $q_0$ represents the most significant qubit. The first $2^{n-1}$ entries of the corresponding state vector would then represent amplitudes for basis states where $q_0$ is $\ket{0}$ and the other entries would represent amplitudes where $q_0$ is $\ket{1}$. This is represented in a decision diagram by a node labeled $q_0$ with two successors labeled $q_1$. By convention, the left (right) successor points to a node that represents  the sub-vector with amplitudes for basis states with $q_0$ assigned $\ket{0}$~($\ket{1}$). This process is repeated recursively until sub-vectors of size 1 (i.e. complex numbers) result. During this process, equivalent sub-vectors are represented by the same node, reducing the overall size of the decision diagram. Furthermore, instead of having distinct terminal nodes for all amplitudes, edge weights are used to store common factors of the amplitudes, leading to even more compaction. Reconstructing the amplitude of a specific state can be done by multiplying the edge weights along the corresponding path. 
In order to increase the readability of the decision diagram edge weights of 1 are omitted. Additionally, nodes with an incoming edge weight of 0 are represented as 0-stubs---indicating that amplitudes of all possible states represented by this part of the decision diagram are zero. 

\begin{figure}[t]
	\centering
	\begin{subfigure}[t]{0.40\linewidth}
		\centering
		\begin{tikzpicture}
		\matrix[matrix of math nodes, left delimiter={[},right delimiter={]}, inner xsep=0] (vector) {
			\frac{1}{\sqrt{2}}\\				
			0\\
			0\\
			\mathbf{\frac{1}{\sqrt{2}}}\\				
		};			
		\begin{scope}[on background layer, black]	
		\node[right=0.6cm of vector-1-1.center] {$\ket{00}$};
		\node[right=0.6cm of vector-2-1.center] {$\ket{01}$};
		\node[right=0.6cm of vector-3-1.center] {$\ket{10}$};
		\node[right=0.6cm of vector-4-1.center] {$\ket{11}$};
		
		\draw[black,-,dashed, very thick,shorten <= -0.6cm] ($(vector-2-1)!0.5!(vector-3-1)$) -- ++(-1.25,0) node[anchor=east] {\(q_0\)};
		
		\draw[black,-,dashed, thick,shorten <= -0.6cm] ($(vector-1-1)!0.5!(vector-2-1)$) -- ++(-1,0) node[anchor=east] {\(q_1\)};
		\draw[black,-,dashed, thick,shorten <= -0.6cm] ($(vector-3-1)!0.5!(vector-4-1)$) -- ++(-1,0) node[anchor=east] {\(q_1\)};
		
		\end{scope}
		\end{tikzpicture}
		\caption{Vector representation}
		\label{fig:statevectorvector}
	\end{subfigure}\hfill
	\begin{subfigure}[t]{0.6\linewidth}
		\centering
		\begin{tikzpicture}	
		\matrix[matrix of nodes,ampersand replacement=\&,every node/.style=vertex,column sep={0.5cm,between origins},row sep={0.9cm,between origins}] (qmdd2) {
			\& \node (m1) {$q_0$}; \& \\
			\node (m2a) {$q_1$}; \& \& \node (m2b) {$q_1$}; \\
			\& \node[terminal] (t3) {1}; \& \\
		};
		
		\draw[thick] ($(m1)+(0,0.5cm)$) -- (m1) node[right, midway]{$\sfrac{1}{\sqrt{2}}$};
		
		\draw (m1.-135) -- ($(m1)!0.5!($(m2a)!0.5!(m2b)$) + (-5mm,0)$) -- (m2a.90);
		\draw[thick] (m1.-45) -- ($(m1)!0.5!($(m2a)!0.5!(m2b)$) + (5mm,0)$) -- (m2b.90);
		
		\draw (m2a.-135) -- ($(m2a.-135) - (1.2mm,2.0mm)$) -- (t3.135);
		\draw (m2a.-45) -- ($(m2a.-45)!0.2!(t3) + (0mm,0)$)  node[zeroterm] {$0$};;
		
		\draw (m2b.-135) -- ($(m2b.-135)!0.2!(t3) + (0mm,0)$)  node[zeroterm] {$0$};;
		\draw[thick] (m2b.-45) -- ($(m2b.-45) - (-1.2mm,2.0mm)$) -- (t3.45);		
		\end{tikzpicture}
		\caption{Decision diagram representation}
		\label{fig:statevectordd}
	\end{subfigure}%
	\vspace*{-2mm}
	\caption{State vector representation}
	\label{fig:statevector}
	\vspace*{-4mm}
\end{figure}
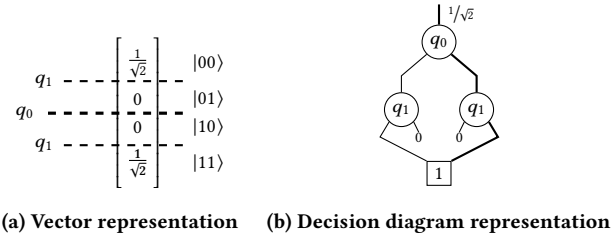

\begin{example}
\label{exp:vectorToDD}
In Fig.~\ref{fig:statevector}, a quantum register is represented both in the vector and decision diagram representation. The annotations of the state vector in Fig.~\ref{fig:statevectorvector} indicates how it is decomposed when the corresponding decision diagram is constructed. Reconstructing the amplitude for a specific state from the decision diagram requires multiplying the edge weights of the corresponding path. For example, reconstructing the amplitude of the state $\ket{11}$ (bold lines in the figure), requires multiplying the edge weight of the root edge (${\sfrac{1}{\sqrt{2}}}$) with the right edge of $q_0$ ($1$) as well as $q_1$ ($1$), i.e.~${\sfrac{1}{\sqrt{2}}} \cdot 1 \cdot 1 = {\sfrac{1}{\sqrt{2}}}$.
\end{example}

Quantum operations are represented in a similar fashion as quantum states. However, due to the square nature of matrices, they are split into four equal sized sub-parts. This is represented in a decision diagram by a node with four successors
the first one representing the sub-matrix in the upper left corner, the second one representing the sub-matrix in the upper right corner, the third one representing the sub-matrix in the lower left corner, and the fourth one representing the sub-matrix in the lower right corner. Apart from this, the decomposition and normalization process is analogue to the one for vectors. 

Similar to array-based simulators, simulation is conducted by multiplying operations onto states. However, due to the different representation, the multiplication must be decomposed with respect to the most significant qubit. Consider again a quantum register $\ket{\phi} = q_0, q_1, \ldots, q_{n-1}$ of $n$ qubits, where $q_0$ represents the most significant qubit, as well as a unitary quantum operation $U$ of size $2^n \times 2^n$. In order to multiply the operation $U$ onto the state~$\ket{\phi}$, they are split into two (in the case of the state vector) and four (in the case of the operation) equally sized parts. Leading to two sub-vectors of size $2^{n-1}$ and four sub-matrices each of size $2^{n-1} \times 2^{n-1}$. 
This represents the modifications of $U$ onto $q_0$ and is represented by a top node labeled $q_0$, with two successor nodes. The splitting process is repeated recursively until vectors of size~$2$ and matrices of size $2 \times 2$ remain. These are multiplied and the resulting new amplitudes are stored into terminal edges. Finally, the edge weights are calculated by extracting common factors of the amplitudes and equivalent sub-vectors are represented by the same node. Hence, multiplication of decision diagrams mainly involves recursive traversals of the involved decision diagrams. On top of that, further optimizations are possible with respect to the precision of the simulation~\cite{DBLP:conf/date/ZulehnerNDW19, niemann2020overcoming}, the \mbox{run-time} performance~\cite{DBLP:conf/date/ZulehnerW19}, or a trade-off of both~\citep{zulehner2020Approx}

\section{Simulation of Decoherence}
\label{sec:sim}

In this work, we aim for the consideration of decoherence errors in the simulation of quantum circuits using decision diagrams. To this end, we provide the motivation and propose an initial approach in this section.
More precisely, this section first reviews decoherence effects that occur in today's quantum computer realizations and afterwards provides a mathematical description of them to be used by simulation approaches in general. Based on that, we then show how those descriptions can be used to conduct such simulations using decision diagrams and discuss how this may affect the performance of corresponding approaches.

\subsection{Qubit Decoherence Errors}
\label{subsec:rep_errors}

Coherence errors occur due to the fragile nature of qubits---leading to the problem that they can only hold information for a limited amount of time. 
More precisely, there are two types of coherence errors that arise~\cite{Tannu2018}:
\begin{itemize}
\item A qubit in a \mbox{high-energy} state~$\ket{1}$ naturally tends to decay to a low energy state~$\ket{0}$, i.e., after a certain amount of time, qubits in a quantum system eventually relax to~$\ket{0}$. This error is called \emph{amplitude damping error} or \emph{T1~error}. 

\item In addition to that, when a qubit interacts with the environment, a phase flip effect might occur. This leads to an error called \emph{phase flip error} or \emph{T2~error}.
\end{itemize}

Recent developments in the physical realization of quantum computers (such as, e.g., in~\cite{Devoret1169, Kelly2018}) show significant improvements in the coherence times and improved possibilities to reduce unwanted interactions of qubits---improving the ``lifetime'' of qubits before decaying to~$\ket{0}$ and reducing the frequency of phase flip errors, respectively. Nevertheless, the underlying errors are still a dominating aspect in all quantum computations and, hence, should also be considered during simulation.%

However, in order to describe them mathematically, the formulation in terms of %
state vectors as reviewed in Section~\ref{sec:bg_quantum} obviously is not sufficient.
Moreover, even a deterministic formulation is not suitable since decoherence heavily relies on probabilistic effects that are not known in advance. 
Hence, a description is needed which incorporates all possible states a quantum system may be in (including the original state but also states resulting from any of the decoherence effects mentioned above with a certain probability). This is accomplished by extending the vector representation introduced in Section~\ref{sec:background} to \emph{density matrices} (also known as \emph{density operators}) as follows:

\begin{definition}
\label{def:density_matrix}
Let $\ket\phi$ be a complex vector representing the state of a quantum system. The corresponding \emph{density matrix} is defined as
\begin{gather}
\begin{aligned}
\rho = \ket{\phi}\bra{\phi} \mbox{ with } \bra{\phi} \coloneqq \ket{\phi}^{\dagger}.\label{eq:mixed_state}
\end{aligned} 
\end{gather}
\end{definition}

\begin{example}
\label{exp:density_matrix}
Consider a system, which is in the state 
\[
\ket{\psi} = \frac{1}{\sqrt{2}} \cdot \ket{00} + 0 \cdot \ket{01} + 0 \cdot \ket{10} + \frac{1}{\sqrt{2}} \cdot \ket{11},
\] 
or, using vector notation, $\sfrac{1}{\sqrt{2}}\cdot\begin{bNiceMatrix}[small] 1&0&0&1\end{bNiceMatrix}^\top$. The corresponding \emph{density matrix} $\rho$ %
is given by
\newcommand\hlight[1]{\tikz[overlay, remember picture,baseline=-\the\dimexpr\fontdimen22\textfont2\relax]\node[rectangle,fill=gray!50,fill opacity = 0.2,draw=gray!50,thick,text opacity =1] {$#1$};} 
\begin{align*}
\begin{bmatrix}
    \frac{1}{\sqrt{2}} \\
    0 \\
    0 \\
    \frac{1}{\sqrt{2}} \\
\end{bmatrix} 
\cdot
\begin{bmatrix}
    \frac{1}{\sqrt{2}} & 0 & 0 & \frac{1}{\sqrt{2}} \\
\end{bmatrix} 
=
\begin{bmatrix}
    \hlight{\frac{1}{2}} & 0 & 0 & \frac{1}{2}  \\
    0 & \hlight{0} & 0 & 0  \\
    0 & 0 & \hlight{0} & 0  \\
    \frac{1}{2} & 0 & 0 & \hlight{\frac{1}{2}}  \\
\end{bmatrix}.
\end{align*}
In contrast to the original vector representation (as illustrated in Example~\ref{exp:matrxi_vector_mul}), the probabilities of measuring specific basic states are now reflected in the diagonal elements (highlighted in gray).
More precisely, the diagonal entries from the first element in the upper-left to the last element in the bottom-right represent the probabilities for measuring $\ket{00}, \ket{01}, \ket{10},$ and $\ket{11}$, respectively. Hence, measuring this state would yield $\ket{00}$ or $\ket{11}$---both with probability of $\sfrac{1}{2}$. %
All other elements of the density matrix represent the coherence in the state. %
\end{example}

Employing this concept allows to %
probabilistically apply decoherence, i.e., the application (or non-application) of decoherence effects by means of probabilities. More precisely, the effects of a decoherence error can be described by \emph{Kraus matrices} defined as follows:

\begin{definition}
\label{def:operator_sum}
Using the operator-sum representation, an error is represented by a tuple $(E_0,E_1, \dots ,E_m)$ of \emph{Kraus matrices} that satisfy the condition
\begin{gather}
  \begin{aligned}
    \sum_{i=0}^m E_i^{\dagger}E_i = I.\label{eq:kraus_condition}
  \end{aligned} 
\end{gather}
Using this notation, the T1 and T2 errors can be represented by~\cite{NC:2000}
\begin{gather}
  \begin{aligned}
    T1 = (E_0, E_1) \mbox{ with } E_0 = \begin{bmatrix} 1 & 0\\ 0 & \sqrt{1-p} \end{bmatrix}, E_1=\begin{bmatrix} 0 & \sqrt{p}\\ 0 & 0 \end{bmatrix}\mbox{and} \label{eq:T1}
  \end{aligned} 
\end{gather}
\begin{gather}
  \begin{aligned}
    T2 = (E_0, E_1) \mbox{ with } E_0=\sqrt{p} \cdot \begin{bmatrix} 1 & 0\\ 0 & 1 \end{bmatrix}, E_1 = \sqrt{1-p} \cdot \begin{bmatrix} 1 & 0\\ 0 & -1 \end{bmatrix}, \label{eq:T2}
  \end{aligned} 
\end{gather}
respectively, where the variable $p$ represents the probability that an error occurs. This probability is a parameter of the specific quantum computer realization (and, hence, needs to be provided by the user).%
\end{definition}

Applying these error descriptions to the current quantum system (represented by a density matrix) can be conducted as follows:

\begin{definition}
\label{def:applying_error}
Applying an error specified by the Kraus matrices $(E_0,E_1, \dots ,E_m)$ to a quantum system given by the density matrix $\rho$ yields the density matrix~\cite{NC:2000}

\begin{gather}
  \begin{aligned}
    \rho^{\prime} = \sum_{i=0}^m E_i \rho E_i^{\dagger}.\label{eq:apply_error}
  \end{aligned} 
\end{gather}
\end{definition}

\begin{example}
\label{exp:applying_noise}
In order to illustrate the concepts above, we apply the amplitude damping (T1) error to the state $\rho$ from Example~\ref{exp:density_matrix}. More precisely, we apply amplitude damping to the second qubit with a probability of 30~\% (p=0.3). The effects of this error are given by the Kraus matrices provided in Eq.~\ref{eq:T1}. Applying each Kraus matrix to the state $\rho$ (as defined in Eq.~\ref{eq:apply_error}) leads to
\newcommand\hlight[1]{\tikz[overlay, remember picture,baseline=-\the\dimexpr\fontdimen22\textfont2\relax]\node[rectangle,fill=gray!50,fill opacity = 0.2,draw=gray!50,thick,text opacity =1] {$#1$};} 
\begin{align*}
\footnotesize
\setlength{\arraycolsep}{4pt}
\medmuskip = 2mu %
\underbrace{\begin{bmatrix}0.5&0&0&0.418\\0&0&0&0\\0&0&0&0\\0.418&0&0&0.35\end{bmatrix}}_{\mathit{E_0 \rho E_0^{\dagger}}} +
\underbrace{\begin{bmatrix}0&0&0&0\\0&0&0&0\\0&0&0.15&0\\0&0&0&0\end{bmatrix}}_{\mathit{E_1 \rho E_1^{\dagger}}} =
\underbrace{\begin{bmatrix}\hlight{0.5}&0&0&0.418\\0&\hlight{0}&0&0\\0&0&\hlight{0.15}&0\\0.418&0&0&\hlight{0.35}\end{bmatrix}}_{\mathit{\rho^{\prime}}}.
\end{align*}
The resulting density matrix accordingly describes the effect of the employed error:
While the probability for measuring $\ket{00}$ remains the same, the probability of measuring $\ket{11}$ has dropped to 35~\% and, additionally, there is now a probability of 15~\% to measure $\ket{10}$. 
In other words, the probability that the second qubit is measured %
$\ket{0}$ has increased by 30~\%---reflecting the damping error assumed above.
\end{example}

Finally, since states are now represented by (density) matrices, %
applying an operation to a state has to be adjusted. Originally, given a quantum state $\ket{\phi}$ and a quantum operation $U$, the corresponding application is conducted by matrix-vector multiplication (as illustrated in Example~\ref{exp:matrxi_vector_mul}). Now, vector states $\ket{\phi}$ are represented by density matrices given by \mbox{$\rho = \ket{\phi}\bra{\phi}$} (cf.~Definition~\ref{def:density_matrix} and Eq.~\ref{eq:mixed_state}). Accordingly, applying $U$ onto $\rho$ is given by

\begin{gather}
  \begin{aligned}
    \rho' = U \rho U^{\dagger}.\label{eq:apply_op_densitymatrix}
  \end{aligned} 
\end{gather}

\subsection{Effect to Simulation}
\label{subsec:effects}

\begin{figure*}[t]
	\begin{subfigure}[b]{0.29\linewidth}
		\centering
		\begin{tikzpicture}
		\matrix[matrix of nodes,ampersand replacement=\&,every node/.style={vertex},column sep={0.5cm,between origins},row sep={1cm,between origins}] (qmdd2) {
			\& \node (m1) {$q_0$}; \& \\
			\node (m2a) {$q_1$}; \& \&\node (m2b) {$q_1$}; \\
			\& \node[terminal] (t) {1}; \& \\
		};
		
		\draw ($(m1)+(0,0.5cm)$) -- (m1) node[right, midway]{$\sfrac{1}{\sqrt{2}}$};	
		
		\draw (m1.-135) -- ($(m1)!0.5!($(m2a)!0.5!(m2b)$) + (-5mm,0)$) -- (m2a.90);
		\draw (m1.-45) -- ($(m1)!0.5!($(m2a)!0.5!(m2b)$) + (5mm,0)$) -- (m2b.90);
		
		\draw (m2a.-135) -- ($(m2a)!0.5!(t) + (-5mm,0)$) -- (t.150);
		\draw (m2a.-45) -- ($(m2a)!0.35!(t) + (1mm,0)$)node[zeroterm] {$0$};;
		
		\draw (m2b.-135) -- ($(m2b)!0.35!(t) + (-1mm,0)$) node[zeroterm] {$0$};;
		\draw (m2b.-45) -- ($(m2b)!0.5!(t) + (5mm,0)$) -- (t.30);;
				
		\end{tikzpicture}
		\caption{Decision diagram of a state vector}
		\label{fig:dd_rep_vector}	
	\end{subfigure}\hfill
	\begin{subfigure}[b]{0.35\linewidth}
		\centering
		\begin{tikzpicture}
		\matrix[matrix of nodes,ampersand replacement=\&,every node/.style={vertex},column sep={0.5cm,between origins},row sep={1cm,between origins}] (qmdd2) {
			\&\&\& \node[draw = none] (top) {}; \&\&\& \\
			\&\&\& \node (m1) {$q_0$}; \&\&\& \\
			\node (m2a) {$q_1$}; \& \&\node (m2b) {$q_1$};  \& \& \node (m2c) {$q_1$}; \& \&\node (m2d) {$q_1$}; \\
			\&\&\& \node[terminal] (t) {1}; \&\&\& \\
		};
		
		\draw (top) -- (m1) node[right,midway] {$\sfrac{1}{2}$};
		
		\draw (m1.-135) -- ($(m1)!0.5!($(m2a)!0.5!(m2a)$) + (-5mm,0)$) -- (m2a.90);
		\draw (m1.-110) -- ($(m1)!0.5!($(m2b)!0.5!(m2b)$) + (-1.5mm,0)$) -- (m2b.90);
		\draw (m1.-70)  -- ($(m1)!0.5!($(m2c)!0.5!(m2c)$) + (1.5mm,0)$) -- (m2c.90);
		\draw (m1.-45)  -- ($(m1)!0.5!($(m2d)!0.5!(m2d)$) + (5mm,0)$) -- (m2d.90);
		
		\draw (m2a.-135) -- ($(m2a)!0.7!(t) + (-13.5mm,0)$) -- (t.170);
		\draw (m2a.-110) -- ($(m2a)!0.35!(t) + (-6mm,0)$) node[zeroterm] {$0$};;
		\draw (m2a.-70) -- ($(m2a)!0.35!(t) + (-4mm,0)$) node[zeroterm] {$0$};;
		\draw (m2a.-45) -- ($(m2a)!0.35!(t) + (-2mm,0)$) node[zeroterm] {$0$};;
		
		\draw (m2b.-135) -- ($(m2b)!0.35!(t) + (-4mm,0)$) node[zeroterm] {$0$};;
		\draw (m2b.-110) -- ($(m2b)!0.7!(t) + (-5mm,0)$) -- (t.150);
		\draw (m2b.-70) -- ($(m2b)!0.35!(t) + (0mm,0)$) node[zeroterm] {$0$};;
		\draw (m2b.-45) -- ($(m2b)!0.35!(t) + (2mm,0)$) node[zeroterm] {$0$};;
		
		\draw (m2c.-135) -- ($(m2c)!0.35!(t) + (-2mm,0)$) node[zeroterm] {$0$};;
		\draw (m2c.-110) -- ($(m2c)!0.35!(t) + (0mm,0)$) node[zeroterm] {$0$};;
		\draw (m2c.-70) -- ($(m2c)!0.7!(t) + (5mm,0)$) -- (t.30);
		\draw (m2c.-45) -- ($(m2c)!0.35!(t) + (4mm,0)$) node[zeroterm] {$0$};;
		
		\draw (m2d.-135) -- ($(m2d)!0.35!(t) + (2mm,0)$) node[zeroterm] {$0$};;
		\draw (m2d.-110) -- ($(m2d)!0.35!(t) + (4mm,0)$) node[zeroterm] {$0$};;
		\draw (m2d.-70) -- ($(m2d)!0.35!(t) + (6mm,0)$) node[zeroterm] {$0$};;
		\draw (m2d.-45) -- ($(m2d)!0.7!(t) + (13.5mm,0)$) -- (t.10);
		\end{tikzpicture}
		\caption{Decision diagram of a density matrix}
		\label{fig:dd_rep_matrix}		
	\end{subfigure}\hfill
		\begin{subfigure}[b]{0.35\linewidth}
		\centering
		\begin{tikzpicture}
		\matrix[matrix of nodes,ampersand replacement=\&,every node/.style={vertex},column sep={0.5cm,between origins},row sep={1cm,between origins}] (qmdd2) {
			\&\&\& \node[draw = none] (top) {}; \&\&\& \\
			\&\&\& \node (m1) {$q_0$}; \&\&\& \\
			\node (m2a) {$q_1$}; \& \&\node (m2b) {$q_1$};  \& \& \node (m2c) {$q_1$}; \& \&\node (m2d) {$q_1$}; \\
			\&\&\& \node[terminal] (t) {1}; \&\&\& \\
		};
		
		\draw (top) -- (m1) node[right,midway] {$\sfrac{1}{2}$};
		
		\draw (m1.-135) -- ($(m1)!0.5!($(m2a)!0.5!(m2a)$) + (-5mm,0)$) -- (m2a.90);
		\draw (m1.-110) -- ($(m1)!0.5!($(m2b)!0.5!(m2b)$) + (-1.5mm,0)$) -- (m2b.90);
		\draw (m1.-70)  -- ($(m1)!0.5!($(m2c)!0.5!(m2c)$) + (1.5mm,0)$) -- (m2c.90);
		\draw (m1.-45)  -- ($(m1)!0.5!($(m2d)!0.5!(m2d)$) + (5mm,0)$) -- (m2d.90);
		
		\draw (m2a.-135) -- ($(m2a)!0.7!(t) + (-13.5mm,0)$) -- (t.170);
		\draw (m2a.-110) -- ($(m2a)!0.35!(t) + (-6mm,0)$) node[zeroterm] {$0$};;
		\draw (m2a.-70) -- ($(m2a)!0.35!(t) + (-4mm,0)$) node[zeroterm] {$0$};;
		\draw (m2a.-45) -- ($(m2a)!0.35!(t) + (-2mm,0)$) node[zeroterm] {$0$};;
		
		\draw (m2b.-135) -- ($(m2b)!0.35!(t) + (-4mm,0)$) node[zeroterm] {$0$};;
		\draw (m2b.-110) -- ($(m2b)!0.7!(t) + (-5mm,0)$) -- (t.150) node [label={[xshift=-8mm, yshift=-0.5mm] $\small 0.836$}] {};
		\draw (m2b.-70) -- ($(m2b)!0.35!(t) + (0mm,0)$) node[zeroterm] {$0$};;
		\draw (m2b.-45) -- ($(m2b)!0.35!(t) + (2mm,0)$) node[zeroterm] {$0$};;
		
		\draw (m2c.-135) -- ($(m2c)!0.35!(t) + (-2mm,0)$) node[zeroterm] {$0$};;
		\draw (m2c.-110) -- ($(m2c)!0.35!(t) + (0mm,0)$) node[zeroterm] {$0$};;
		\draw (m2c.-70) -- ($(m2c)!0.7!(t) + (5mm,0)$) -- (t.30) node [label={[xshift=1mm, yshift=-0.5mm] $\small 0.836$}] {};
		\draw (m2c.-45) -- ($(m2c)!0.35!(t) + (4mm,0)$) node[zeroterm] {$0$};;
		
		\draw (m2d.-135) -- ($(m2d)!0.35!(t) + (2mm,0)$) node[zeroterm] {$0$};;
		\draw (m2d.-110) -- ($(m2d)!0.6!(t) + (7mm,0)$) -- (t.20) node [label={[xshift=8mm, yshift=0.5mm] $\small 0.3$}] {};
		\draw (m2d.-70) -- ($(m2d)!0.35!(t) + (6mm,0)$) node[zeroterm] {$0$};;
		\draw (m2d.-45) -- ($(m2d)!0.7!(t) + (13.5mm,0)$) -- (t.10) node [label={[xshift=14mm, yshift=0.7mm] $\small 0.7$}] {};
		\end{tikzpicture}
		\caption{State after applying T1 error}
		\label{fig:dd_after_to_noise}
	\end{subfigure}	
	\vspace*{-2mm}
	\caption{Decision diagram representation of states}
	\label{fig:dd_rep_examples}
	\vspace*{-4mm}
\end{figure*}
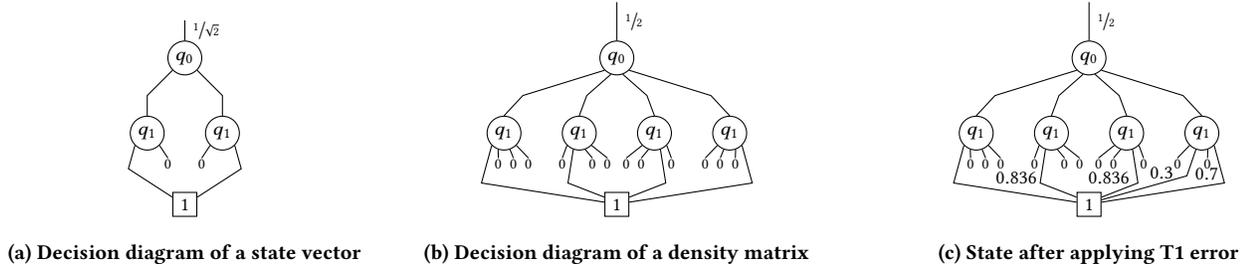

Having the mathematical description of decoherence errors as reviewed above, simulation approaches can accordingly be extended.
This is particularly straightforward for array-based approaches as proposed in~\cite{DBLP:journals/corr/WeckerS14,DBLP:journals/corr/SmelyanskiySA16,qxSimulator2017,qiskit,Steiger2018projectqopensource,CirqPythonFramework,atos2016} and reviewed in Section~\ref{sec:array_based_sim}. 
Here, the major challenge of considering decoherence is ``only'' given by the fact that the state is represented by density matrices rather than vectors.
This can simply be addressed by extending the corresponding data structures (i.e., the arrays). The required operations (in particular, for multiplying and adding matrices)
are supported by the corresponding libraries anyway and can also be extended in a straightforward fashion. Also, improvements employed by parallel executions can readily be utilized. Because of that,
implementations of array-based simulation approaches supporting the consideration of decoherence are already available (see, e.g.,~\cite{qiskit,atos2016,qxSimulator2017,DBLP:journals/corr/WeckerS14,CirqPythonFramework,jones2018quest,DBLP:journals/corr/SmelyanskiySA16}).

For approaches based on decision diagrams, the consideration of decoherence errors, however, may pose much more severe challenges (which have not be considered in the literature thus far). First, it remains open how severely the consideration of decoherence errors (and the corresponding need of representing states and operations through density matrices) harms the ability of decision diagrams to represent states and operations in a compact fashion. As reviewed in Section~\ref{subsec:dd_simulation}, being able to represent certain quantum states and quantum operations in a more compact fashion constitutes one of the main advantages of simulation using decision diagrams.
The question remains whether this compactness can still be maintained when decoherence errors are additionally considered. 
To this end, consider the following example:

\begin{example}
Recall the state considered before in Example~\ref{exp:density_matrix} in both the vector as well as in the density matrix representation. %
The corresponding decision diagram representations %
are provided in Fig.~\ref{fig:dd_rep_vector} and Fig.~\ref{fig:dd_rep_matrix}, respectively.
As can be seen, the size of the decision diagrams are rather similar (4 nodes vs.~6 nodes, although the corresponding vector/density matrix have a size of~$2^2$ and~$2^{2}\times 2^{2}$). %
This, at least, shows that considering decoherence errors does not necessarily harm the ability of decision diagrams to represent states/operations in a compact fashion.\footnote{Note that this does not mean that decision diagrams \emph{always} provide a compact representation for a quantum state/operation. In fact, previous work~\cite{zulehner2017advanced} clearly showed that the worst-case complexity of decision diagrams is exponential---even though polynomial representations are possible for certain applications. The example, however, shows that this characteristic is not completely lost when considering decoherence errors.}
Moreover, also more detailed evaluations (summarized later in Section~\ref{sec:eval}) confirm this observation.
\end{example}

Considering the example (and the evaluations summarized in Section~\ref{sec:eval}) suggests that decision diagrams might remain compact in many cases---even if decoherence errors are considered. 
But having a compact representation alone is not sufficient. Also efficient realizations of the corresponding matrix-matrix operations (most notably multiplication and addition), as reviewed in Section~\ref{subsec:rep_errors}, are required. For multiplication, related work such as~\cite{zulehner2017advanced} already provides efficient solutions since matrix-matrix multiplication already is a core operation of simulation without decoherence. Addition, however, which is required for applying the error effect as defined in Eq.~\ref{eq:apply_error}, has not been that frequently required thus far and turns out to be particularly challenging. 

More precisely, recall that adding two matrices is done by adding all elements sharing the same index. Hence, this requires access to \emph{all} single matrix elements---something for which a decision diagram has to be \emph{completely} traversed. This is in stark contrast with multiplication, where addition is required as well, but the operands are often just sub-parts of the involved decision diagrams (which substantially reduces the size of the decision diagrams that need to be traversed).
Additionally, decision diagram representations of operations are of very sparse nature, so that during the multiplication often one part of the addition equals zero.

Overall, this suggests that decision diagrams also show promise for quantum circuit simulation when considering decoherence errors (something which has not been considered thus far). At the same time, it also unveils challenges which remains to be addressed, namely how to efficiently %
apply the decoherence to the state without having to traverse the entire decision diagram. 

\section{Advanced Simulation Approach}
\label{sec:opt}

In order to address the shortcomings unveiled above, we investigated how the required operations (particularly Eq.~\ref{eq:apply_error}) can efficiently be realized on decision diagrams. A major obstacle is that applying decoherence effects heavily relies on the addition of matrices which requires access to \emph{all} single matrix elements and, in turn, triggers the complete traversal of the decision diagrams---causing an exponential overhead independently of how compact they can be represented.

Accordingly, we looked for alternatives that either completely avoid the addition of matrices or, at least, only conduct it on (smaller) sub-matrices. %
Our investigations eventually resulted in such alternatives whose main idea rests on the following three observations:

First, %
adding two matrices is not always necessary. In particular, the T2 error can be described by multiplications only. In fact, w.l.o.g., the effect of this error on a single qubit can be described by 
\begin{align*}
\begin{bmatrix}
    a & b\\
	c & d
\end{bmatrix} 
\longmapsto 
\begin{bmatrix}
    a & (2p-1)b\\
	(2p-1)c & d
\end{bmatrix}
,
\end{align*}
where $p$ represents the probability. That is, two of the elements are not changed at all, while the remaining ones are just multiplied by $2p-1$. Hence, applying the T2 error can be reduced to two multiplications---without the need for any addition. %

Second, applying the error as specified in Eq.~\ref{eq:apply_error} is inefficient and allows to apply only one error effect to one qubit at a time. By explicitly enforcing the error effects directly on the corresponding nodes of the affected qubits, we can apply all desired effects to all qubits with just one traversal of the decision diagram.

Third, if the matrix is represented in terms of decision diagrams, it is not always necessary to do the addition on the entire matrix.
In fact, in decision diagrams, every qubit is represented by one or more nodes. Applying a decoherence error to a qubit actually only modifies the outgoing edges from the nodes representing this specific qubit. 
All predecessor nodes are only indirectly affected when the edge weights are normalized (as described in Section~\ref{subsec:dd_simulation}).

Based on those observations, we are proposing an advanced method of applying this operation on decision diagrams:
Instead of doing the matrix-matrix multiplications followed by the addition separately, we apply the decoherence effects in one step---exploiting our observation that applying an operation to a qubit only modifies the outgoing edge weights of nodes representing it. Decoherence can therefore accordingly be applied by directly modifying all those edges. In doing so, we also gain full control over how the operation is applied---allowing us to exploit the first two observations as well.

\begin{example}
\label{exp:directly_applying_noise}
The advanced approach is illustrated by reusing Example~\ref{exp:applying_noise}. Recall that, in this example, we apply an amplitude damping~(T1) decoherence error with a probability of 30~\% to $q_1$ of the quantum state $\rho$ from Example~\ref{exp:density_matrix}. In contrast to the earlier example, we now apply the error directly to the decision diagram of $\rho$ (given in Fig.~\ref{fig:dd_rep_matrix}). 
To this end, we apply the effects of the T1 error to all nodes labeled~$q_1$ leading to the following transformations 
\begin{align*}
\begin{bmatrix}
    a & b\\
	c & d
\end{bmatrix} 
\longmapsto
\begin{bmatrix}
    a+dp & b\sqrt{1-p}\\
	c\sqrt{1-p} & d-dp
\end{bmatrix}.
\end{align*}
As defined in Section~\ref{subsec:dd_simulation}, the four outgoing edges of a node in a decision diagram represent the matrix elements a, b, c and d from left to right. Hence, modifying the fourth node labeled $q_1$ with probability $p=0.3$ leads to
\begin{align*}
\begin{bmatrix}
    0 & 0\\
	0 & 1
\end{bmatrix} 
\longmapsto
\begin{bmatrix}
    0 + 1 \cdot 0.3 & 0 \\
	0 & 1 - 1 \cdot 0.3 \\
\end{bmatrix}
=
\begin{bmatrix}
    0.3 & 0\\
	0 & 0.7
\end{bmatrix} 
\end{align*}
The other nodes labeled $q_1$ are modified in the same way, leading to the new decision diagram shown in Figure~\ref{fig:dd_after_to_noise}. This new decision diagram represents $\rho$ after the T1 decoherence error has been applied and is equal to $\rho^{\prime}$ from Example~\ref{exp:applying_noise}.
\end{example}

\section{Evaluation}
\label{sec:eval}

\begin{table*}[ht]
		\vspace{.75cm}
	\caption{Experimental results}
	\label{tab:results}
	\setlength{\tabcolsep}{8pt}
	\centering	
	\begin{tabular}{r|rr|rr|rr|rr}
		& \multicolumn{4}{c|}{Array-based} & \multicolumn{4}{c}{Decision diagram-based} \\
		& \multicolumn{2}{c|}{QLM} & \multicolumn{2}{c|}{Qiskit} & \multicolumn{2}{c|}{Without Improvement} & \multicolumn{2}{c}{With Improvement} \\
		\#Qubits & Mem\,[MB] & Time\,[s] & Mem\,[MB] & Time\,[s] & Mem\,[MB] & Time\,[s] & Mem\,[MB] & Time\,[s] \\\hline\hline
		\begin{filecontents*}{csv/results_AD.csv}
Qubits,Time,Mem,Time,Mem,Time,Mem,Time,Mem
10,142.11,2.06,261.57,40.04,94.44,0.47,41.94,0.16
11,190.21,4.29,313.65,97.27,95.03,0.87,41.00,0.16
12,390.43,13.54,505.40,174.69,97.62,2.40,39.97,0.17
13,1166.96,58.22,1297.79,366.26,98.22,7.47,40.48,0.17
14,4241.86,265.40,4287.89,1114.23,100.58,22.49,40.74,0.17
15,16535.59,1217.52,,,103.10,72.89,43.21,0.17
16,,,,,107.69,269.04,41.70,0.17
17,,,,,109.55,1067.78,94.46,0.27
18,,,,,,,97.21,0.27
,,,,,,,\vdots,\vdots
36,,,,,,,129.85,528.62
37,,,,,,,135.08,1255.23
38,,,,,,,137.02,2816.20
\end{filecontents*}
\csvreader[
		late after line=\\,
		late after last line=\\\cline{2-9},
		]{csv/results_AD.csv}	
		{1=\Name, 2=\Ngates, 3=\TimeQL, 4=\MemLi, 5=\TimeQX, 6=\MemQX, 7=\TimeQP, 8=\MemQP, 9=\Time}
		{\Name  & \Ngates & \TimeQL & \MemLi & \TimeQX & \MemQX & \TimeQP & \MemQP & \Time}		
		\end{tabular}
\end{table*}

In this section, we summarize the core results of our evaluations conducted in order to investigate the effect of considering decoherence errors in the simulation of quantum circuits using decision diagrams.
To this end, we took the state-of-the-art decision \mbox{diagram-based} simulator from~\cite{zulehner2019package,zulehner2017advanced} and extended this implementation to additionally support decoherence errors. This led to one version in which error support has been added in a straightforward fashion (i.e., directly applying the concepts described in Section~\ref{subsec:rep_errors}) and another version in which error support has been added in an advanced fashion (i.e., applying the methods described in Section~\ref{sec:opt}). In addition to that and for the purpose of comparison, we also considered two state-of-the-art array-based simulators, namely  \emph{Linalg} from the commercial \emph{Atos Quantum Learning Machine} (QLM)~\cite{atos2016} as well as the \emph{density\_matrix} simulator from IBM's Qiskit~\cite{qiskit}. 

For the evaluations, we assumed the amplitude damping (T1) error with 0.2~\% probability and the phase flip (T2) error with 0.1~\% probability (applied each time a qubit has been used). 
As quantum algorithm, we considered the \emph{Quantum Fourier Transform}~(QFT~\cite{NC:2000}) %
with an increasing number of qubits. %
This is an ideal choice for the purpose of this evaluation, since QFT (1)~constitutes a core element of numerous existing quantum applications (such as Shor's factorization method~\cite{DBLP:journals/siamcomp/Shor97}, quantum phase estimation~\cite{NC:2000} and the hidden subgroup problem~\cite{NC:2000}), 
(2)~is an established benchmark for evaluating the effects of decoherence errors in related work (e.g.,~\cite{Chaudhary2019}), and
(3)~additionally has the benefit that, thus far, it represents one of the best cases for simulation using decision diagrams without considering decoherence (showing a linear scalability in memory and \mbox{run-time}  with respect to the number of qubits). Therefore, the effects of considering decoherence errors can ideally be investigated using QFT.

All evaluations have been conducted on a system using 5 cores running at a clock frequency of 2.2~GHz and 1.5~TB of RAM. While the QLM simulator runs directly on this system, we ported the Qiskit simulator and the simulators using decision diagrams to this machine utilizing Docker~\cite{Merkel:2014:DLL:2600239.2600241}. We choose Docker since its virtualization overhead is negligible~\cite{Felter2015}. By this, all simulators have been evaluated using the same hardware resources.

Table~\ref{tab:results} summarizes the obtained results. The first column provides the number of considered qubits. In the remaining columns, we list the peak memory usage in MB as well as the total simulation time in real time seconds for all considered approaches, i.e., Atos' and IBM's array-based QLM and Qiskit, respectively, as well as the considered decision diagrams-based approach with and without the improvement from Section~\ref{sec:opt}. Note that the accumulated CPU time of the array-based approaches would be substantially larger than the real time values listed in 
Table~\ref{tab:results}, since both approaches heavily utilize concurrency enabled by the available five cores during the simulation. Cells without any entries represent instances where the timeout of one hour has been exceeded.

The results clearly confirm the observations from above: Considering decoherence errors during simulation does not necessarily harm the compact representation of decision diagrams.
In fact, the respective memory requirements for the decision diagrams remain rather moderate (never more than 150~MB), while they sky-rocket for the array-based approaches (certainly, the main reason why those approaches only scale up to 15 or 14 qubits, respectively).
At the same time, it also can be seen that \mbox{run-time}  becomes much more the limiting factor for the decision \mbox{diagram-based} simulation. This confirms the discussion from Section~\ref{subsec:effects}
and shows the impact of the advanced method of conducting the required operations as described in Section~\ref{sec:opt}---eventually yielding improvements of several orders of magnitude and more than twice the scalability.

\section{Conclusion}
\label{sec:conclusion}
Decision diagrams provide a promising alternative for quantum circuit simulation due to their capability of representing vectors and matrices in a much more compact fashion than, e.g., array-based methods. But no work existed yet which 
investigated whether these promising effects can be maintained when decoherence errors---still, a dominating aspect in quantum computing---are additionally assumed.
This work sheds light on this. We observed that considering decoherence errors not necessarily harms the compact representation, but leads to new challenges 
with respect to conducting the respectively required operations in a \mbox{run-time-efficient} fashion. In order to address these challenges an advanced method has been proposed, which mitigates the negative effects and led to improvements of several orders of magnitudes. By this, we showed that quantum circuit simulation using decision diagrams remains a promising approach also when decoherence errors are considered.

\begin{acks}
This work has partially been supported by the University of Applied Sciences PhD program of the State of Upper Austria (managed by the FFG), by the LIT Secure and Correct Systems Lab funded by the State of Upper Austria, as well as by the BMK, BMDW, and the State of Upper Austria in the frame of the COMET program (managed by the FFG).
\end{acks}
\newpage

\bibliographystyle{ACM-Reference-Format}
\bibliography{IEEEabrv,../../bib/lit_quantum,../../bib/lit_header,../../bib/lit_misc,../../bib/lit_mymisc,../../bib/lit_others,../../bib/lit_othersrev,../../bib/lit_rev,../../bib/lit_adiabatic,../../bib/lit_memristor,../lit_simulation,../new-references}

\end{document}